\documentclass[prb,preprint]{revtex4-1} 
\usepackage{amsmath}  
\usepackage{amsfonts} 
\usepackage{graphicx} 

\begin{document}


\title{Measurement of Gravitational Time Dilation: \\An Undergraduate Research Project}

\author{M.\ Shane Burns}
\email{sburns@coloradocollege.edu}
\author{Michael D.\ Leveille}
\author{Armand R.\ Dominguez}
\author{Brian B.\ Gebhard}
\author{Samuel E.\ Huestis}
\author{Jeffery Steele}
\affiliation{Department of Physics, Colorado College, Colorado Springs, CO 80903}

\author{Brian Patterson}
\email{Brian.Patterson@usafa.edu}
\author{Jerry F.\ Sell}
\author{Mario Serna}
\author{M.\ Alina Gearba}
\author{Robert Olesen}
\author{Patrick O'Shea}
\author{Jonathan Schiller}
\affiliation{Department of Physics, U.S. Air Force Academy, Colorado Springs, Colorado 80840}


\date{\today}

\begin{abstract}
General relativity predicts that clocks run more slowly near massive objects. The effect is small---a clock at sea level lags behind one 1000 m above sea level by only 9.4 ns/day.  Here, we demonstrate that a measurement of this effect can be done by undergraduate students. Our paper describes an experiment conducted by undergraduate researchers at Colorado College and the United States Air Force Academy to measure gravitational time dilation. The measurement was done by comparing the signals generated by a GPS frequency standard (sea-level time) to a Cs-beam frequency standard at seven different altitudes above sea level.  We found that our measurements are consistent with the predictions of general relativity. 
\end{abstract}

\maketitle 

\section{Introduction} 
General relativity predicts that clocks tick more slowly near massive objects. This well-understood and well-tested effect is referred to as gravitational time dilation (GTD). Most undergraduate physics majors are aware of the effect. It was even depicted recently in the major motion picture \emph{Interstellar}. Physics students who are interested in the effect may study it theoretically, but are seldom able to experimentally test it because it requires a very precise time measurement. Einstein discussed GTD in detail in his 1916 paper.\cite{einstein1916} The fact that his prediction wasn't tested directly until 1959 by Pound and Rebka\cite{pound1959} illustrates the difficulty of actually doing the measurement. 

The effect is, however, important. Although it was a required consideration for the engineers that designed the global positioning system (GPS) over 30 years ago,\cite{ashby2002, ashby2003} it is still relatively difficult to test experimentally using equipment typically found in an undergraduate lab. In late 2014 the United States Air Force Academy (USAFA) acquired four cesium-beam frequency standards from surplus created by a restructuring of other governmental laboratories. In the spring of 2015 faculty members at USAFA and Colorado College (CC) began a collaboration with students at both institutions. The goal of the faculty was to help the students design and execute an experiment to measure GTD. With faculty assistance, the students designed the experiment, wrote the data acquisition software, analyzed the data, and contributed to the writing of this paper.

In this paper we describe our experiment to directly test GTD by using a cesium-beam frequency standard and a GPS receiver. Each instrument contains a 10 MHz quartz oscillator whose frequency is subtly adjusted to match an underlying physical reference. The reference for the Cs instrument is set by the frequency of the transition between two energy levels of $^{137}$Cs atoms inside the instrument itself whereas the GPS receiver relies on signals from the orbiting GPS satellites. Although the GPS satellites are in Earth orbit, the frequency signals they generate are corrected to mimic the behavior of frequency standards at the Earth's geoid near sea level. Thus, if the altitude of our two instruments is changed, only the Cs frequency standard is affected by GTD.

During our experiment we placed a Cs frequency standard and GPS receiver at several different altitudes and at various locations in Colorado. This allowed us to measure GTD as a function of distance from Earth's geoid. The geoid is a gravitational equipotential surface corresponding approximately to mean sea level. Its actual shape is irregular due to variations in the mass distribution of Earth.\cite{fowler}  This paper will provide details of the experimental setup and procedure so that other students and teachers can reproduce the result with a modest financial investment. In fact, similar ventures have also been reported by amateur clock enthusiasts.\cite{vanbaak2007} 

In order to measure the GTD effect at a given altitude, we monitored the phase shift between the 10-MHz signal from a Cs frequency standard and the 10-MHz signal from a GPS frequency standard. By convention, the GPS frequency is corrected to the frequency at Earth's geoid near sea level\cite{ashby2003}. The geoid is defined in the World Geodetic System WGS84 standard\cite{ngawebsite} on which GPS coordinates are referenced. By being above the Earth's GPS reference geoid, the Cs frequency standard suffers less time dilation than the GPS standard and hence the phase shift between the two signals increases as time goes on. In our experiment we monitored this shift for several days at several different altitudes ranging from 1339 m to 4288 m above the geoid. 

\section{Theoretical Background}
\label{sec:theory}
The Schwarzschild metric\cite{hartel} describes the spacetime around a spherically symmetric object of mass $M$ as
\begin{equation}
\label{eq:schwarzschild}
-ds^2 = c^2 d\tau ^2 = \left(c\, dt \sqrt{1-\frac{R_s}{r}}\right)^2 - \left(\frac{dr}{\sqrt{1-\frac{R_s}{r}}}\right)^2 - r^2 d\theta^2 - r^2 \sin^2\theta d\phi^2
\end{equation}
where $R_s = \frac{2GM}{c^2}$ is the Schwarzschild radius. The spacetime interval is $ds$ and the proper time $d\tau$ is the time read on a clock that travels along the spacetime interval $ds$. For the Earth, the Schwarzschild coordinate $r$ is very nearly the radial distance from the Earth's center.\cite{radial} The coordinates $\theta$ and $\phi$ are the usual angular coordinates of a spherical polar coordinate system. The coordinate time $t$ is the time read on a clock far from the mass.

The Schwarzschild radius for the Earth is $R_s = 8.9$ mm, so $R_s/r \ll 1$ for points above the Earth's geoid. Hence, we can approximate the metric above the Earth's geoid as
\begin{equation}
\label{eq:weak}
-ds^2 = c^2 d\tau ^2 = \left(1-\frac{R_s}{2r}\right)^2 c^2\, dt^2 - \left(1+\frac{R_s}{2r}\right)^2 dr^2 - r^2 d\theta^2 - r^2 \sin^2\theta d\phi^2.
\end{equation}
We can use this metric to compute the proper time interval $d\tau$ measured by a clock at a distance $r$ from the center of the Earth, and at rest with respect to the surface. If we choose to align the $z$-axis with the Earth's axis of rotation, then $d\theta = 0$ and $dr = 0$, hence, 
\begin{equation}
\label{eq:delta_tau}
\left(\frac{d\tau}{dt}\right)^2 = \left(1-\frac{R_s}{2r}\right)^2 - \left(\frac{r}{c} \frac{d\phi}{dt}\right)^2 \sin^2\theta.
\end{equation}
The first term on the right hand side of Eq.~(\ref{eq:delta_tau}) is the GTD effect. The second term is the special relativistic time dilation effect due to the fact that clocks on the Earth's surface are traveling at a speed $r\frac{d\phi}{dt}$ relative to observers at rest far from the Earth's surface.

Letting $r = R + h$, where $R$ is the distance from the center of the Earth to the geoid and $h$ is the distance of the clock above the geoid, we can rewrite Eq.~(\ref{eq:delta_tau}) as
\begin{equation}
\label{eq:delta_tau_h_exact}
\left(\frac{d\tau}{dt}\right)^2 = \left[1-\frac{R_s}{2R} \left(1+\frac{h}{R}\right)^{-1}\right]^2 - \left(\frac{\omega R}{c}\right)^2 \sin^2\theta \left(1+\frac{h}{R}\right)^{2},
\end{equation}
where $\omega = 7.29\times10^{-5}\, \mathrm{s}^{-1}$ is the Earth's sidereal rotation rate. For our experiment, the maximum value for $h = 4300$ m so $h/R \lesssim 7\times10^{-4} \ll 1$ so we can approximate Eq.~(\ref{eq:delta_tau_h_exact}) as
\begin{equation}
\label{eq:delta_tau_h_1}
\left(\frac{d\tau}{dt}\right)^2 = \left[1-\frac{R_s}{2R} \left(1-\frac{h}{R}\right)\right]^2 - \left(\frac{\omega R}{c}\right)^2 \sin^2\theta \left(1+2\frac{h}{R}\right).
\end{equation}
The factors $R_s/(2R) = 6.95 \times 10^{-10}$ and $(\omega R/c)^2 = 2.41 \times 10^{-12}$ are also small, so taking the square root of both sides of Eq.~(\ref{eq:delta_tau_h_1}) and dropping terms of order $(R_s/(2R))^2$ gives
\begin{equation}
\label{eq:delta_tau_h}
\frac{d\tau}{dt} =  1-\left(\frac{R_s}{2R}+ \frac{\omega^2 R^2}{2 c^2}\sin^2\theta\right) + \left( \frac{R_s}{2R} - \frac{\omega^2 R^2}{c^2}\sin^2\theta\right) \frac{h}{R}.
\end{equation}
We can simplify this further by noting that $(\omega R/c)^2 \ll R_s/(2R)$. Physically this means that  special relativistic time dilation is small compared to GTD for our clocks. Ignoring the special relativistic effect gives us a relatively simple equation relating $d\tau$, the time interval read on a clock on the Earth's surface at a height $h$ above the geoid, and $dt$, the coordinate time interval read on distant clocks,
\begin{equation}
\label{eq:dtau_GTD}
d\tau = \left[ 1-\frac{R_s}{2R} +  \frac{R_s}{2R}  \left(\frac{h}{R}\right) \right]dt.
\end{equation}
The proper time, $\tau$, elapsed on a clock at a height $h$ above the geoid is, therefore, less than the coordinate time $t$. 

In our experiment we measure the elapsed time difference $\Delta \tau \equiv \tau_h - \tau_0$, where $\tau_h$ is the time elapsed on a clock a height $h$ above the geoid and $\tau_0$ is the time elapsed on a clock at the geoid. The time given from a GPS system is the time at the World Geodetic System (WGS84) reference ellipsoid which is very close to the geoid.\cite{ashby2003} Using Eq.~(\ref{eq:dtau_GTD}) we find that
\begin{equation}
\label{eq:Delta_tau_1}
\Delta \tau = \frac{R_s}{2R^2} h\, t= \frac{GM}{R^2} \frac{1}{c^2}h\, t = \frac{g}{c^2}h\, t.
\end{equation}
Note that $\frac{GM}{R^2}$ is just the acceleration due to gravity, $g$, at the geoid. Also, the coordinate time $t$ differs from the proper time $\tau_h$ by only about a part in a million over several days, so we can write Eq.~\ref{eq:Delta_tau_1} as
\begin{equation}
\label{eq:Delta_tau}
\Delta\tau = \frac{g}{c^2} h\, \tau_h.
\end{equation}
The rate at which the time difference between a clock at height $h$ and a clock on the geoid increases is
\begin{equation}
\label{eq:rate}
\frac{\Delta\tau}{\tau_h} = \frac{g}{c^2} h.
\end{equation}

In our experiment we measured $\Delta \tau / \tau_h$ at several different altitudes. The theory above predicts that if we plot $\Delta \tau / \tau_h$ versus $h$ we should get a straight line with a slope $g/c^2$. Using the WGS84 values for $M$ and $R$ gives $g = 9.7983\, \mathrm{m\, s}^{-2}$ and $g/c^2 = 1.0902 \times 10^{-16}\, \mathrm{m}^{-1}$. We can convert $g/c^2$ to more useful units by multiplying by the number of ns per day to obtain
\begin{equation}
\label{eq:prediction}
\frac{g}{c^2} = 9.4194\,  \frac{\mathrm{ns}}{\mathrm{day\, km}}.
\end{equation} 
Thus a clock used to time a full rotation of the earth will measure the day to be approximately an extra 10 ns/day longer for every km of altitude above the reference geoid. 

\section{Experimental Setup}

\subsection{Equipment}
\label{sec:equipment}
Our team started with four HP 5071A Primary Frequency Standards (Cs clocks) acquired by the Air Force Academy. We selected the three clocks having the most stable output frequencies to use for this experiment. Three Trimble ThunderBolt GPS Disciplined Clocks generated the 10-MHz signal which represented `sea-level time'. 
For each setup, the phase differences between the GPS clock signal and the Cs clock signal was measured with an Agilent 53000 Series Frequency Counter. This is a time interval counter to report the time difference between the upward zero-crossings of the two input 10-MHz signals.  The output of the frequency counter's time interval measurement was recorded in a text file on a computer. Table~\ref{tbl:components} lists the major components for each setup and summarizes their functions.

Three almost identical setups were used in order to measure GTD in multiple locations and to provide redundancy in data collection. We designated the three Cs clocks as clock A, B, and D. Colorado College used clock A and B. USAFA used clock D. The primary difference between the CC and USAFA setups was the data acquisition software. The CC software was written in Python, ran on MacBook Pro computers, and was connected to the frequency counters using a LAN connection. 
The USAFA software was written with LabVIEW and connected to a Windows computer using the frequency counter's GPIB interface.
\begin{table}[ht!]
\centering
\caption{Major components of the apparatus used to measure GTD.}
\begin{ruledtabular}
\begin{tabular}{l p{8cm}}
Device & Notes  \\
\hline	
HP 5071A Primary Frequency Standard  & The 10-MHz output signal from this frequency 
									   standard is referred to as the `Cs clock signal' 
									   or $\tau_h$.  \\
Trimble ThunderBolt GPS Disciplined Clock  & The GPS signal is used to produce a 10-MHz 
											 frequency standard. The output 
											 signal is referred to as the `GPS signal' 
											 or $\tau_0$.  \\
Agilent 53000 series Frequency Counter  	& Measures the time interval, 
									  $\Delta \tau \equiv \tau_h - \tau_0$, between 
									  GPS signal and clock signal.   \\
Data acquisition computer  			& Interfaces with the HP 5071A and the GPS receiver. It also 
									  records the time of the measurement and the output
									  of the frequency counter ($\Delta \tau$) to a text file.
\end{tabular}
\end{ruledtabular}
\label{tbl:components}
\end{table}
 
The HP 5071A Cs clock works by using electronic transitions between the two hyperfine ground states of Cs-133 atoms to control the output frequency of a slaved oscillator. Each Cs clock used in this experiment was configured to output a 10-MHz signal. Clocks A and B were prepared for data acquisition by running a Python script that sent commands over a USB/RS-232 serial cable to the HP 5071A.  The script set the output frequency and the current time (UTC) and date on the HP 5071A. Clock D was prepared for operation manually using the front-panel keypad on the HP 5071A. 

The Trimble ThunderBolt GPS-Disciplined-Clock system consists of a GPS antenna and receiver electronics. The ThunderBolt clock works by using the GPS signal to discipline a temperature-stabilized quartz oscillator to generate a 10-MHz signal. We controlled the GPS clock by using the ThunderBolt Monitor Program (Tboltmon) running on the data acquisition computer and connected by a USB/RS-232 serial connection.  

The Agilent 53000 Series Frequency Counters have two input channels. The Cs clock signal was connected to one channel and the GPS clock signal was connected to the other. The frequency counter triggers the start of a timer when one of the signals crosses upward above a particular voltage, and then stops the timer when the second signal crosses the same threshold.  This measured time interval, $\Delta \tau\equiv \tau_h - \tau_0$, is sent to the computer over a LAN or GPIB connection where it is recorded in a text file along with a time-stamp from the computer. 

\begin{figure}[ht!] 
\centering
\includegraphics[width=6.25in]{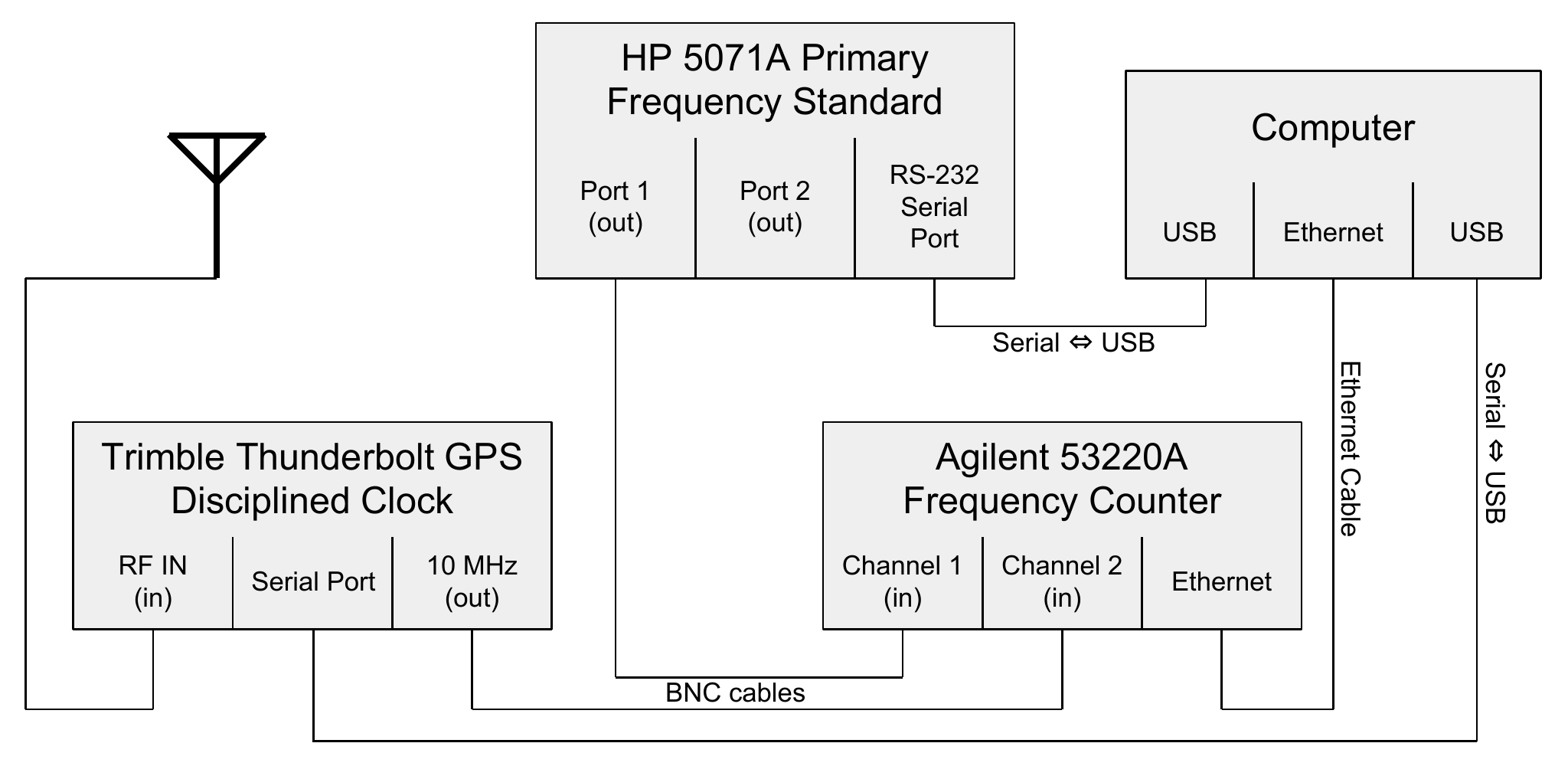}
\caption{Schematic of the data collection system.}
\label{fig:setup}
\end{figure}

The schematic in Figure~\ref{fig:setup} shows the details of how the apparatus was typically set up. The GPS signal from the antenna is connected to the GPS disciplined clock. The GPS clock's 10-MHz signal is connected to Channel 2 of the frequency counter using a BNC cable. The Cs clock signal from Port 1 of the 5071A is connected to Channel 1 of the frequency counter. These two signals are compared and the time interval, $\Delta \tau $, is measured as described above. The time interval is sent to the data acquisition computer over a LAN connection where it is recorded along with the time-stamp.

\subsection{Data Collection}
Twenty-two different data collection runs were done at seven different locations across Colorado at altitudes ranging from 1340~m (Trainor Ranch near La Junta, CO) to 4288~m (the summit of Pikes Peak). The shortest data collection run was done at Arapahoe Basin Ski Patrol Headquarters and lasted less than 24 hours. It ended early when a ski area employee inadvertently shut down the data acquisition computer. Aside from this one run, all of our data runs lasted at least three days. The longest runs, done on the summit of Pikes Peak, were about 2 weeks. Table~\ref{tbl:data_runs} lists the location, altitude, and equipment used for each  data collection run. The altitudes listed in the table were recorded from the GPS units.   
\begin{table}[ht!]
\centering
\caption{Summary of data collection runs.}
\begin{ruledtabular}
\begin{tabular}{l l c c}
Start Date  & Location & Altitude (m) & Cs Clock \\
\hline	
17/03/2016 & US Air Force Academy& 2165 & D \\
05/04/2016 & Colorado College & 1848 & B \\
25/04/2016 & Trainor Ranch & 1340 & B \\
25/04/2016 & Trainor Ranch & 1338 & A \\
27/04/2016 & Colorado College Cabin & 2683 & B \\
28/04/2016 & Colorado College & 1848 & A \\
01/05/2016 & Arapahoe Basin Resort & 3294 & A \\
01/05/2016 & Arapahoe Basin Patrol HQ & 3785 & B \\
05/06/2016 & US Air Force Academy& 2165 & D \\
09/06/2016 & Colorado College & 1846 & B \\
17/06/2016 & US Air Force Academy& 2165 & D \\
22/06/2016 & US Air Force Academy& 2165 & D \\
28/06/2016 & US Air Force Academy& 2165 & D \\
08/07/2016 & Colorado College & 1846 & B \\
12/07/2016 & Colorado College & 1845 & B \\
18/07/2016 & Colorado College & 1843 & B \\
26/07/2016 & Colorado College & 1844 & B \\
30/08/2016 & Colorado College & 1846 & B \\
07/09/2016 & US Army Pikes Peak Research Lab & 4288 & B \\
07/09/2016 & US Army Pikes Peak Research Lab & 4288 & D \\
15/09/2016 & US Army Pikes Peak Research Lab & 4288 & B \\
15/09/2016 & US Army Pikes Peak Research Lab & 4288 & D \\
\end{tabular}
\end{ruledtabular}
\label{tbl:data_runs}
\end{table}

\section{Analysis}

\subsection{Correction for data wrapping} 
\begin{figure}[ht!] 
\centering
\includegraphics{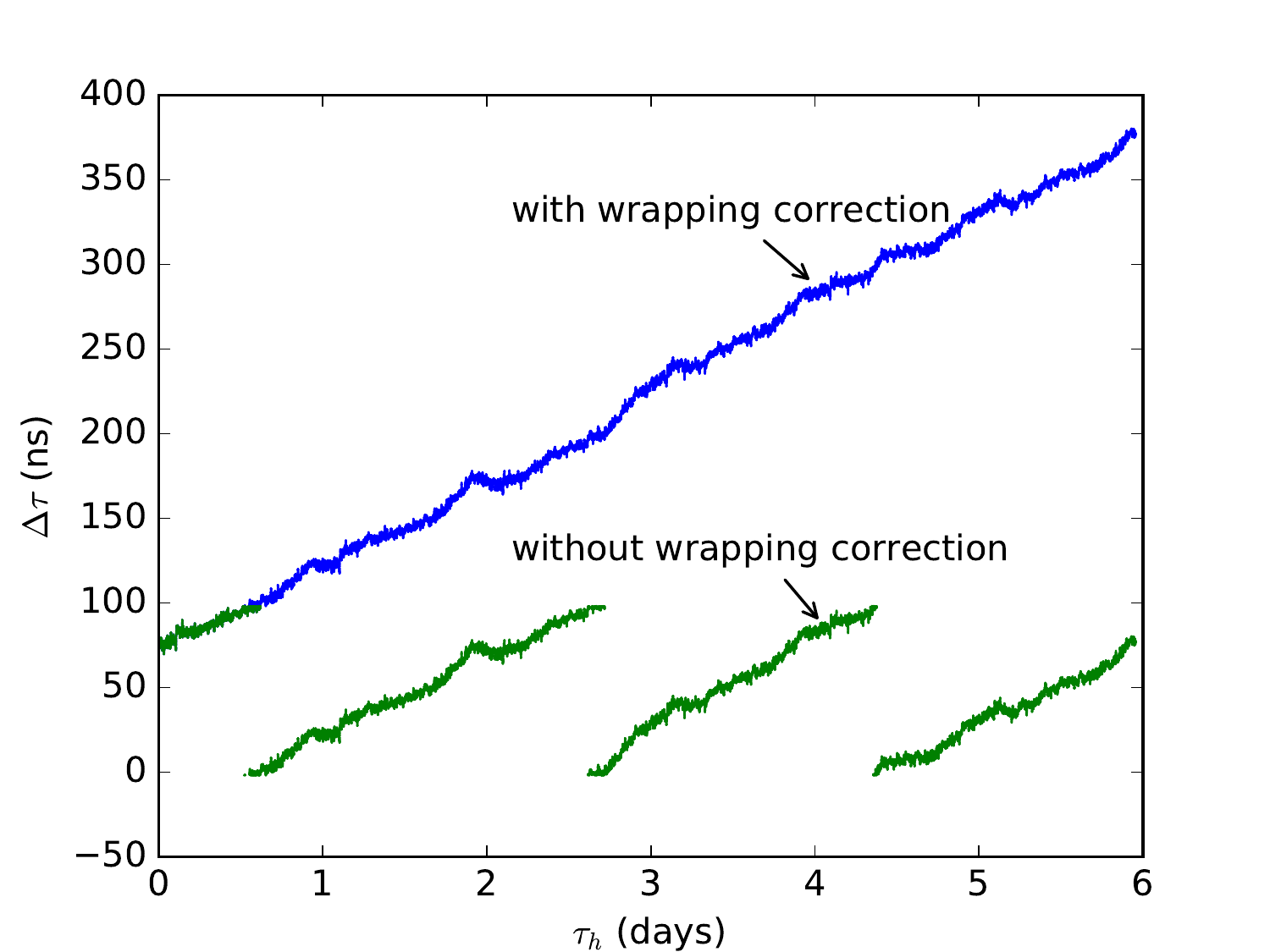}
\caption{Plot of time interval, $\Delta \tau$, between the GPS and Cs-beam signals showing the data before and after applying the data wrapping correction.}
\label{fig:wrapped}
\end{figure}

Section~\ref{sec:equipment} describes how the frequency counter measures a time interval between the signals from the GPS and the Cs-beam frequency standard. As a result of GTD, the GPS frequency is slightly larger than the frequency of the Cs-beam standard. This, in turn, causes the time interval between the two signals to continually increase. When the time interval reaches 100 ns, one of the signals is a full signal period behind the other and the time interval registered by the counter is zero. The time interval then begins to grow until again reaching 100 ns and the process repeats. The counter effectively `wraps' the time interval back into a value between 0 and 100 ns. The first step in our data analysis was to add back the missing 100 ns intervals. This was accomplished using a Python script that scanned consecutive time intervals for a 100-ns jump and, when found, added 100 ns to all the successive intervals. Figure~\ref{fig:wrapped} is a plot of $\Delta \tau$ versus $\tau_h$ for a run done on Pikes Peak. The plot shows data before and after the data-wrapping correction.

\subsection{Individual data run analysis}
In Section~\ref{sec:theory} we derived Eq.~\ref{eq:Delta_tau_1} which predicts that $\Delta \tau$ should increase linearly with time with a slope equal to $gh/c^2$. We fit the data from each run to the function $\Delta \tau = a\, \tau_h + b$. The slope, $a$, should be equal to $gh/c^2$. The intercept, $b$, is just the arbitrary time difference between the two signals when the experiment starts.

\begin{figure}[ht!] 
\centering
\includegraphics{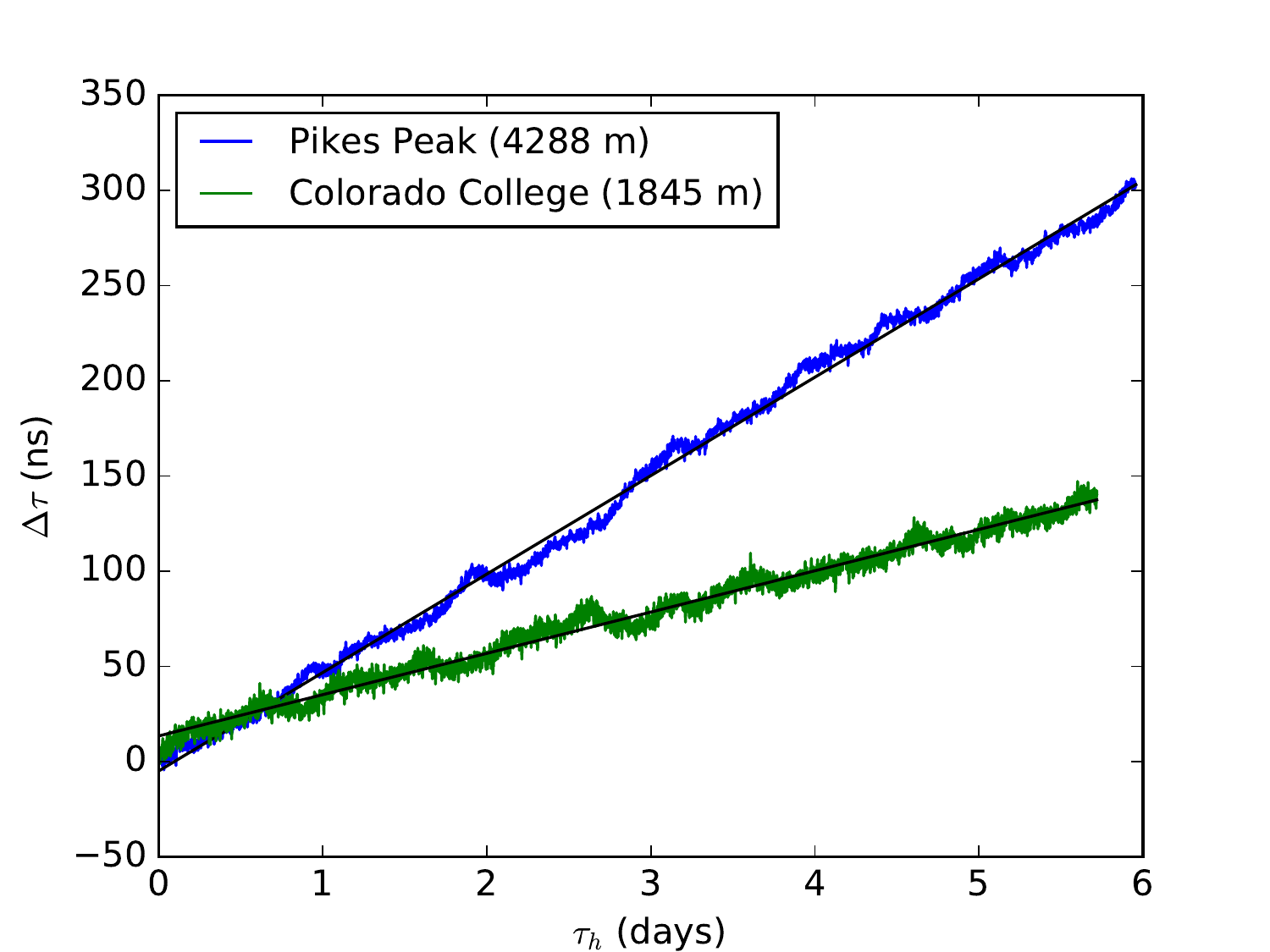}
\caption{Plot of the time difference $\Delta \tau$ as a function of time for data runs done on the summit of Pikes Peak and at Colorado College.}
\label{fig:sample_plot}
\end{figure}

Figure~\ref{fig:sample_plot} shows a plot of the measured time difference, $\Delta \tau$, as a function of time for two data collection runs along with their linear fits. One run was done on Pikes Peak at an altitude of 4288~m and one was done at Colorado College at 1845~m. The graphs show that the time difference $\Delta \tau$ does indeed increase linearly with $\tau_h$. The plots also show that the slope is greater at higher altitudes as predicted by general relativity. The fits gave slopes of 51.7~ns/day at 4288~m and 21.7~ns/day at 1845~m. The predicted values are 40.4~ns/day and 17.4~ns/day, respectively.

The measured slope values are larger than the predictions of GTD for both altitudes. We suspect that this difference is due to bias in the Cs frequency standard. Such systematic biases are well documented.\cite{shirley2001} The 5071A data sheet\cite{5017Adatasheet} specifies that the long term stability of the standard is less than about $8.5 \times 10^{-14}$ over a period of five days. This translates to a bias of approximately 7 ns/day. The magnitude of this bias was confirmed by moving all three Cs clocks to the same location and then measuring the time difference between clocks over a period of about five days (see the appendix). Our attempt to mitigated this bias is discussed below. 

\subsection{Analysis of altitude dependence}
Eq.~\ref{eq:rate} shows that the time difference between a clock at $h$ and a clock on the geoid increases linearly with altitude. In order to test this prediction we fit individual data runs for each clock to a linear function 
\begin{equation}
\label{eq:altitude_fit}
\frac{\Delta \tau}{\tau_h} = \alpha h + \beta. 
\end{equation}
According to GTD theory, the slope $\alpha$ should be equal to $g/c^2 = 9.4194\, \mathrm{ns}\, \mathrm{day}^{-1}\, \mathrm{km}^{-1}$ (Eq.~\ref{eq:prediction}).

The clocks used in the experiment were transported to locations with different altitudes to measure the GTD dependence on altitude. This required us to shutdown and restart them each time they were moved. Each time the clocks were restarted, the frequency output of each was different than the nominal output frequency of 10 MHz. In all cases the difference was consistent with the manufacturer's accuracy specification. Clock B and D were restarted several times at the same altitude to estimate the size of this effect. We determined that the fractional deviation in the frequency of clock B was $3.2 \times 10^{-14}$  or 2.8 ns/day. The fractional deviation for clock D was $5.9 \times 10^{-14}$ or 5.1 ns/day. Both of these deviations are consistent with the manufacturer specifications for the frequency standards.\cite{5017Adatasheet} We also found that when the clocks were restarted clock B would consistently produce a higher frequency than the other clocks. This was an extremely small effect and smaller than the manufacturer's specified deviation. We did, however, try to measure the effect by comparing the clocks. The appendix summarizes the results of that measurement. In order to correct for this small systematic effect, we included an intercept $\beta$ in the fits. 
\begin{figure}[ht!]
\centering
\includegraphics{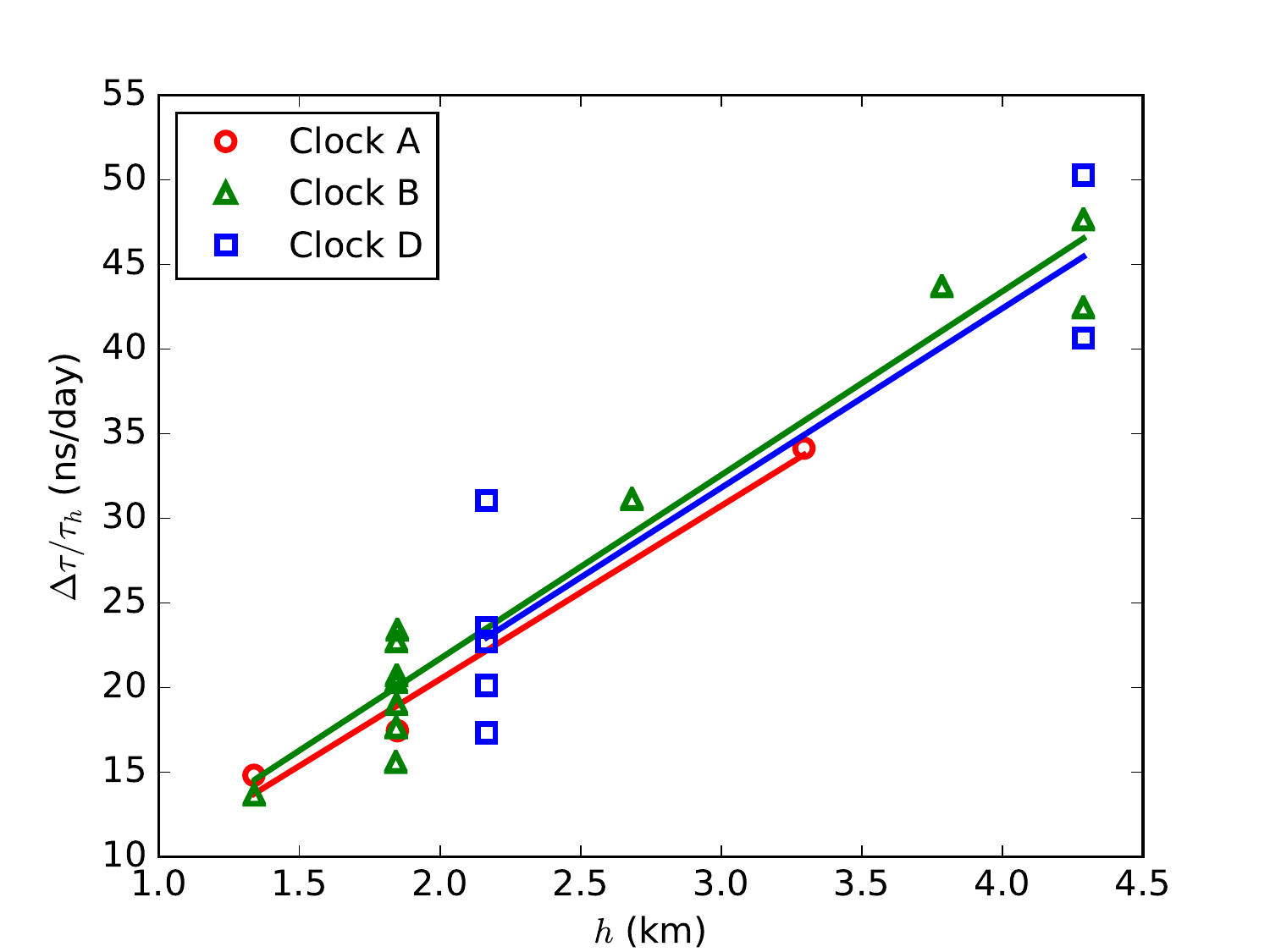}
\caption{Plot of the time dilation rate, $\Delta \tau/\tau_h$, versus height, $h$, above the geoid. We subtracted the intercept derived from the fit for each frequency standard from each data point and the fit in order to correct for different systematic biases for each frequency standard.}
\label{fig:rate_vs_altitude}
\end{figure}

The time dilation rates, $\Delta \tau/\tau_h$, for the three frequency standards are shown in Fig.~\ref{fig:rate_vs_altitude} as a function of height $h$, along with unweighted linear fits to these data. The plots show that once we have corrected the data for the systematic biases by subtracting the intercepts derived from the fits, the three frequency standards give essentially the same result. The results of the fits for all three clocks are summarized in Table~\ref{tbl:results}. For all but one of the frequency standards the value of the intercept is consistent with zero. The measured values of $\alpha$ for all three Cs frequency standards are larger than the theoretical predictions, but the discrepancy isn't significant for any of the measurements. Frequency standards A and D give results within about half a standard deviation of the theoretical prediction and frequency standard B give a result just under two standard deviations from theoretical prediction. 
\begin{table}[ht!]
\centering
\caption{Summary of fits of altitude dependence for each frequency standard.}
\begin{ruledtabular}
\begin{tabular}{c c c c}
Cs Clock 	& $\alpha$ ($\mathrm{ns}\, /\mathrm{day}/\, \mathrm{km}$) & $\frac{\alpha - \alpha_\mathrm{theory}}{\sigma}$	& $\beta$ ($\mathrm{ns}/\, \mathrm{day}$)  \\
\hline	
A 					& $10.2 \pm 1.3$ 										  & 0.60			& $-0.9 \pm 3.1$ \\
B 					& $10.85 \pm 0.78$ 										  & 1.83			& $4.1 \pm 2.1$ \\
D 					& $10.6 \pm 2.2$ 										  & 0.54			& $4.0 \pm 6.4$ 
\end{tabular}
\end{ruledtabular}
\label{tbl:results}
\end{table}   

\section{Conclusion}
The effects of general relativity are well known and excite undergraduate students, but are difficult to demonstrate. This experiment shows that a measurement of GTD is indeed possible by comparing the signal from a GPS time standard and a Cs frequency standard. 

In the experiment described here we demonstrated the GTD effect by comparing the GPS standard to a Cs frequency standard (see Figure~\ref{fig:sample_plot}). We also demonstrated the GTD altitude dependence. The experiment was done using three different Cs frequency standards. In all cases we obtained results consistent with the prediction of general relativity (see Figure~\ref{fig:rate_vs_altitude} and Table~\ref{tbl:results}).  
 
\appendix*   
\section{Time Standard Comparison}
\label{sec:comparison}
\begin{figure}[ht!] 
\centering
\includegraphics[width=4.5in]{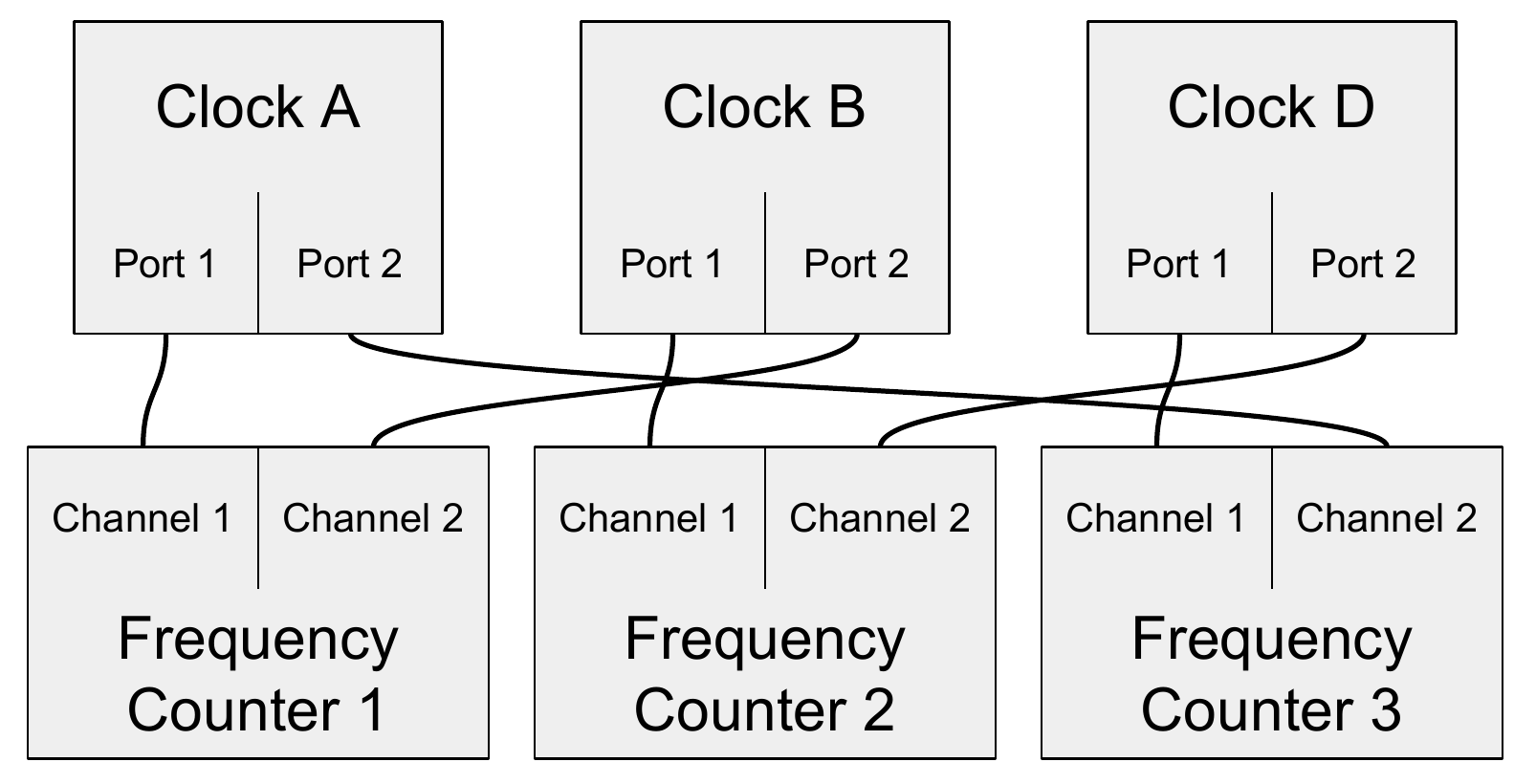}
\caption{Experimental setup for measuring the drift rate between the three clocks.}
\label{fig:calibration}
\end{figure}

We performed one experiment to explore the bias in frequency of all three Cs clocks and several to measure the bias between clocks A and B. We accomplished the three-clock run by measuring the phase drift between the three clocks over a period of 4.5 days.  The Cs clocks were connected as shown in the schematic diagram in Figure~\ref{fig:calibration}. Each Cs clock has two 10-MHz output ports. The frequency counters were connected to the computer via a LAN connection. We used the same software that we used to measure the GPS clock and Cs clock time differences to record the time differences between each of the three clocks.  

\begin{figure}[ht!] 
\centering
\includegraphics[width=4.5in]{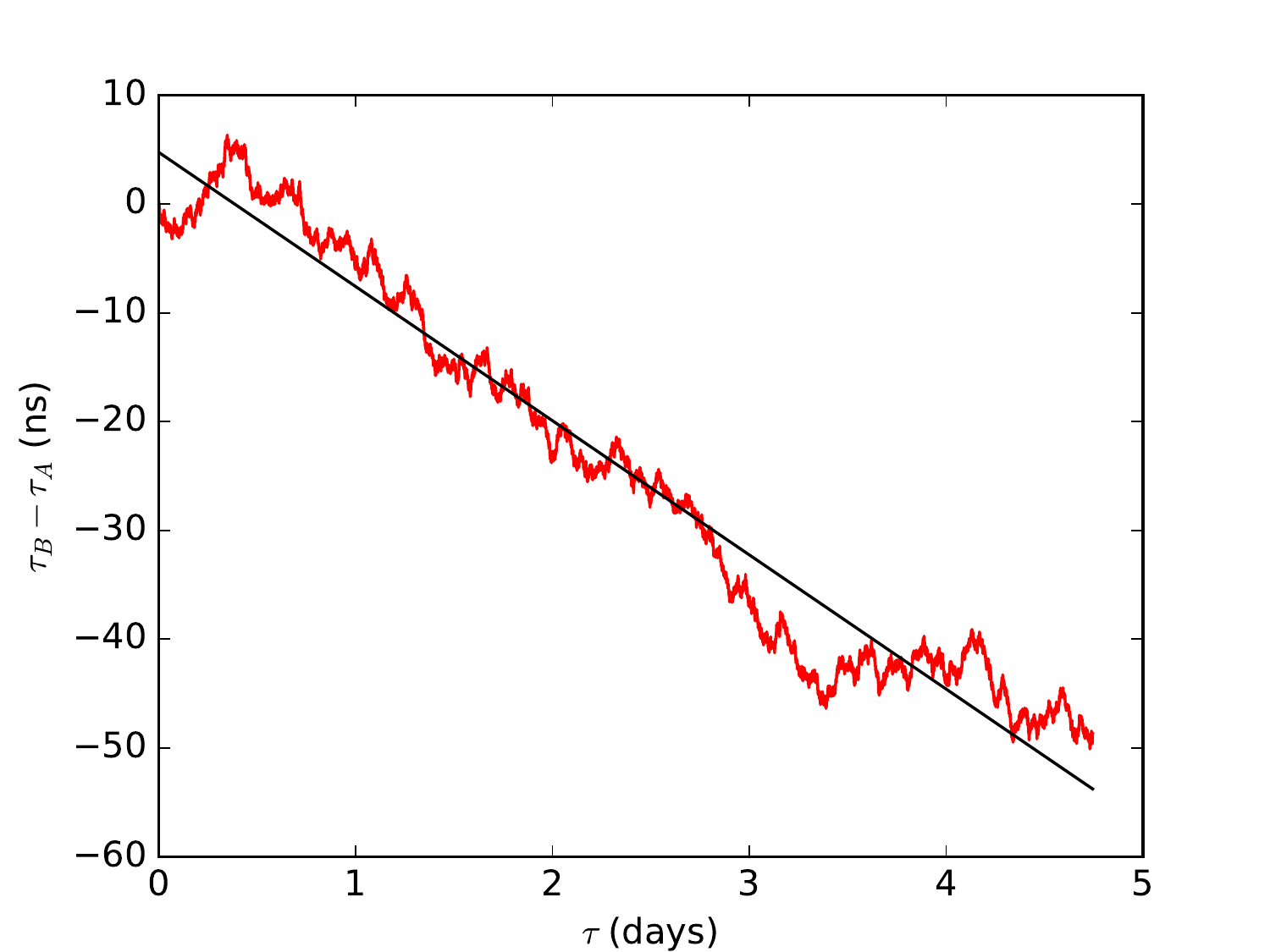}
\caption{The phase drift between Cs clocks A and B from the three-clock run.}
\label{fig:BAdrift}
\end{figure}

We found that there was a systematic drift between all three clocks. Figure~\ref{fig:BAdrift} shows the drift between Cs clocks A and B. 
A linear fit gives a systematic drift rate between Cs clocks A and B, $(\tau_B - \tau_A)/\tau_B = -12$~ns/day. We were able to estimate the variability of this drift from the other clock A and B runs. We found that the rms deviation in the drift rates between Cs clocks A and B was 4 ns/day.  

The measurements for the other Cs clocks gave similar results.  We found that $(\tau_D - \tau_B)/\tau_D = 5$~ns/day  and  $(\tau_A - \tau_D)/\tau_A = 7$~ns/day.  If we assume  the same uncertainty for these  measurements as the measurement of Cs clocks A and B, the results are consistent with the  biases determined from the intercepts of the  fits  summarized in  Table~\ref{tbl:results}.

\begin{acknowledgments}

We gratefully acknowledge the support of Brian McGarvey and the RX research group at Fort Meade for providing access to the Cs atomic clocks. We also acknowledge the helpful assistance from personnel at A-Basin and Trainor Ranch. Special thanks go to John Bristow for arranging the use of the facilities at the United States Army Pikes Peak Research Laboratory. This work was funded by grants from the Mellon Foundation for civilian/military collaboration and by the Colorado College Center for Immersive Learning and Engaged Teaching. J.F.S. acknowledges support from the Air Force Office of Scientific Research and the National Science Foundation (Grant No. 1531107).

\end{acknowledgments}


\begin{thebibliography}{99}
\bibitem{radial} The Schwarzschild coordinate $r = (A/4\pi)^{1/2}$ where $A$ is the area of a sphere centered on the spherical mass. However, $r$ is very close to the distance from the center of the Earth for the weak gravitational curvature around the Earth.
\end{thebibliography}
\end{document}